\documentclass[12pt]{iopart}
\usepackage{latexsym}

\def\cR{{\cal R}}
\def\Tr{{\rm Tr}}
\def\d={\buildrel \rm def \over =}
\def\ket#1{\mid~\!\!\!{#1}~\!\!\rangle}
\def\bra#1{\langle~\!\!{#1}~\!\!\!\mid}

\begin{document}\jl{1}

\title[\bf Range-dimension conditions
for quantum correlations]{\bf Necessary
and sufficient range-dimension
conditions for bipartite quantum
correlations}
\author{F Herbut\footnote[1]{E-mail:
fedorh@infosky.net}}
\address{Faculty of Physics, University of
Belgrade, POB 368, Belgrade 11001,
Yugoslavia and Serbian Academy of
Sciences and Arts, Knez Mihajlova 35,
11000 Belgrade}

\date{\today}

\begin{abstract}
As it is well known, every mixed or
pure state of a bipartite quantum
system is given by a statistical
operator, which determines, in terms of
its two reduced statistical operators,
the subsystem states. Necessary and
sufficient conditions for the existence
of a composite-system state, and,
separately, for the possibility of its
being correlated or uncorrelated in
terms of the range dimensions of the
three mentioned statistical operators
are derived. As a corollary, it is
shown that it cannot happen that two of
the mentioned dimensions are finite and
the third is infinite.
\end{abstract}

\maketitle

\normalsize \rm

It is assumed throughout that one can
speak of quantum correlations in a
composite $1+2$ quantum system if and
only if a correlated composite-system
statistical operator $\rho_3$ is given.
(The index $3$ instead of $12$ is used
for reasons that will become clear
below in condition (1).) Let, further,
$$ \rho_1\equiv \Tr_2\rho_3,\qquad
\rho_2\equiv \Tr_1\rho_3 $$ be the
reduced statistical operators
(physically: the states of the
subsystems), where "$\Tr_2$" and
"$\Tr_1$" denote the respective partial
traces. Let the dimensions of the
ranges $\cR(\rho_i)$, i. e., the {\it
range dimensions}, be denoted by $d_i$,
$i=1,2,3$.

Physically, a statistical operator
$\rho_3$ is a general, i. e., a pure or
mixed, state of a composite system. It
is {\it uncorrelated} if
$\rho_3=\rho_1\otimes \rho_2$, and it
is {\it correlated} otherwise.

The theorem to be proved in this note
works with three range-dimension
conditions:

{\it The cyclic inequality conditions}:
$$ d_1\leq d_2d_3,\qquad d_2\leq
d_3d_1, \qquad d_3\leq
d_1d_2;\eqno{(1)}$$

{\it the common lower bound condition}:
$$ 2\leq d_1,d_2;\eqno{(2)}$$

and, finally,

{\it the product condition}: $$
d_3=d_1d_2.\eqno{(3)}$$

It is easily seen that (3) implies (1)
and obviously the former does not
follow from the latter, i. e., (3) is a
{\it stronger requirement} than (1).\\

The {\it theorem} on {\it
range-dimension conditions} for
bipartite states goes as follows:

(i) For every (correlated or
uncorrelated) state $\rho_3$ conditions
(1) are valid; and {\it vice versa},
for every three natural numbers
$d_1,d_2,d_3$ satisfying condition (1)
there exists at least one state
$\rho_3$ implying them as its range
dimensions.

(ii) One can have a {\it correlated}
composite-system state $\rho_3$ if and
only if, in addition to the cyclic
inequality conditions (1), also the
common lower bound condition (2) is
valid.

(iii) For every {\it uncorrelated}
state $\rho_3$ condition (3) is valid;
and {\it vice versa}, if three natural
numbers $d_1,d_2,d_3$ satisfy condition
(3), then there exists at least one
uncorrelated state $\rho_3$ for which
these numbers are the range
dimensions.\\

At first sight one may be puzzled
because "correlated" and "uncorrelated"
are mutually exclusive concepts for
states, and the corresponding claimed
conditions do not exclude each other:
both conditions (ii) and (iii) can be
simultaneously valid.

The answer, of course, lies in the fact
that in the mentioned case both
correlated and uncorrelated states
exist. Namely, the "sufficiency" in the
condition does not claim that the
condition necessarily implies
correlations or lack of them
respectively; it only implies the
existence of a correlated (or an
uncorrelated) state with the given
range dimensions.

It should be pointed out that none of
the three dimensions is assumed to be
finite.

Conditions (1) will be seen to follow
from a remarkable fact: $$ d_3=1\qquad
\Rightarrow \qquad d_2=d_1,
\eqno{(4)}$$ which is known \cite{HV},
but, perhaps, not well known. It is
easy to see that condition (4) follows
from (1), but the latter has a wider
scope than the former: it covers all
states, not only the pure ones
($d_3=1$).

It is a {\it corollary} of conditions
(1) that {\it one cannot have precisely
one of the three dimensions infinite}.
If, e. g., $d_1$ were infinite, and
$d_2$ and $d_3$ were finite, this would
contradict the first inequality in (1).
The symmetrical arguments hold for the
other two cases. The conditions
obviously allow all three of the
dimensions or any two or none to be
infinite.\\

The theorem is a modest contribution to
the study of quantum correlations, and
the latter are important for quantum
information theory, as well as for
quantum communication and quantum
computation theories \cite{Vedral}.\\

The rest of this note is devoted to
{\it a proof of the theorem}. We begin
by proving claim (i).

 {\it To prove neccesity} of the first
condition in (1), we assume that an
arbitrary composite-system statistical
operator $\rho_3$ is given. Every such
operator has a purely discrete (finite
or infinite) spectrum (\cite{Sim},
theorems VI.16. and VI.21.). Hence, we
can write it in spectral form: $$
\rho_3=\sum_{n=1}^{d_3}r_n\ket{\Psi^{(n)}}_3\bra
{\Psi^{(n)}}_3.\eqno{(5)}$$

On account of (4), one has $$
d_1^{(n)}=d_2^{(n)},\qquad n=1,2,\dots
,d_3, \eqno{(6)}$$ where the symbols
$d_i^{(n)}$ denote the respective
dimensions of the ranges
$\cR(\rho_i^{(n)})$, and $\rho_i^{(n)}$
are the reduced statistical operators
of the pure characteristic states
$\ket{\Psi^{(n)}}_3$ in (5),
$i=1,2;\quad n=1,2,\dots ,d_3$.

Taking the partial trace over subsystem
$2$ in (5), one obtains $$
\rho_1=\sum_{n=1}^{d_3}r_n\rho_1^{(n)}.
\eqno{(7)}$$

We replace each $\rho_1^{(n)}$ in (7)
by a spectral decomposition into pure
states with positive characteristic
values, i. e., we write $$
\rho_1=\sum_{n=1}^{d_3}r_n
\sum_{j=1}^{d_1^{(n)}}r_j^{(n)}
\ket{\phi_j^{(n)}}_1\bra{\phi_j^{(n)}}_1.
\eqno{(8)}$$ Let us be reminded of the
known fact that, in general, the state
vectors corresponding to the pure
states of which the state is a mixture
span (as linear combinations and
limiting points) the topological
closure $\bar \cR(\rho)$ of the range
of the statistical operator $\rho$ that
corresponds to the mixture. Hence, one
can conclude that $$ \bar
\cR(\rho_1)=\sum_{n=1}^{d_3}\bar
\cR(\rho_1^{(n)}) \eqno{(9)}$$ is
valid. (The sum in (9) is an ordinary
sum of subspaces, i. e., the LHS is the
linear and topological span of the
union of the RHS subspaces. The terms
need not be linearly independent, let
alone orthogonal.)

As to the dimensions, (9) evidently
implies $$ d_1^{(n)}\leq d_1\leq
\sum_{n'=1}^{d_3}d_1^{(n')},\qquad
n=1,2,\dots ,d_3.\eqno{(10)}$$ (The
second equality is achieved if the sum
in (9) is a direct one or, in
particular, an orthogonal one.)
Naturally, also the symmetrical
inequalities hold true (and they are
proved by the symmetrical argument): $$
d_2^{(n)}\leq d_2\leq
\sum_{n'=1}^{d_3}d_2^{(n')},\qquad
n=1,2,\dots ,d_3.\eqno{(11)}$$

Substituting (6) in the second
inequality in (10), one obtains $$
d_1\leq \sum_{n=1}^{d_3}d_2^{(n)}.$$
Utilizing the first inequality in (11),
one, further, has $$ d_1\leq
\sum_{n=1}^{d_3}d_2=d_2d_3.$$

The second inequality in (1),  i. e.,
$d_2\leq d_3d_1$, is proved
symmetrically. The last relation in
(1), i. e., $d_3\leq d_1d_2$, is known
(see, e. g. \cite{FY}, relation (11)
there).\\

 {\it To prove sufficiency} of the
three inequalities in (1), we assume
first that $d_3$ is the dominant
quantity, i. e., that we have $$
d_2\leq d_1\leq d_3,\eqno{(12)}$$ and
we give a construction of a statistical
operator $\rho_3$ having the given
dimensions. (Within the case of
dominance of $d_3$, the other
possibility, namely $d_1\leq d_2$ is
handled symmetrically.)

Further, we take an orthonormal (ON)
basis $\{\ket{\phi^{(i)}}_1:i=1,2,\dots
,d_1\}$ spanning the range
$\cR(\rho_1)$, and an ON basis
$\{\ket{\chi^{(j)}}_2:j=1,2,\dots
,d_2\}$ spanning the range
$\cR(\rho_2)$. (If a range is infinite
dimensional, then it is understood that
an ON basis spans the topological
closure of the range.) Proceeding
further, we make the direct products:
$\bar \rho_3^{(i,j)}\equiv
\ket{\phi^{(i)}}_1\bra{\phi^{(i)}}_1
\otimes
\ket{\chi^{(j)}}_2\bra{\chi^{(j)}}_2$,
for all pairs $(i,j)$.

Next, we form the sequence $$
\rho_3^{(n=1)}\equiv \bar
\rho_3^{(1,1)},\enskip
\rho_3^{(n=2)}\equiv \bar
\rho_3^{(2,2)}, \enskip \dots ,\enskip
\rho_3^{(n=d_2)}\equiv \bar
\rho_3^{(d_2,d_2)};$$
$$\rho_3^{(n=d_2+1)}\equiv \bar
\rho_3^{(d_2+1,1)},\enskip
\rho_3^{(n=d_2+2)}\equiv \bar
\rho_3^{(d_2+2,1)},\enskip \dots,
\enskip \rho_3^{(n=d_1)}\equiv \bar
\rho_3^{(d_1,1)},$$ join to it any
$(d_3-d_1)$ states $\bar
\rho_3^{(i,j)}$ from the rest as $$
\rho_3^{(n=d_1+1)},\enskip
\dots,\enskip \rho_3^{(n=d_3)}.$$

 Since $d_3\leq d_1d_2$ (the third
inequality in (1)), there is a
sufficient number of $\bar
\rho_3^{(i,j)}$ states for this. If
$d_2=d_1$, then the second row is, of
course, omitted. Analogously, if
$d_1=d_3$, the last subset of states is
omitted. Finally, we take a
decomposition of $1$ into $d_3$
positive numbers $w_n$:
$1=\sum_{n=1}^{d_3}w_n$, and the
constructed $\rho_3$ is by definition
the mixture $$\rho_3\equiv
\sum_{n=1}^{d_3}w_n\rho_3^{(n)}.$$ It
is easy to see that this state, which
is a mixture of orthogonal pure states,
has the dimensions $d_i,\quad i=1,2,3$
given at the beginning of our
sufficiency proof for (1).\\

To proceed with our proof of
sufficiency of the three inequalities
in  condition (1), for the existence of
a composite-system statistical operator
$\rho_3$ with the given  dimensions, we
assume now that $d_3$ is not the
dominant quantity, i. e., that we have
$$ d_2,d_3\leq d_1.\eqno{(13)}$$
(Again, the subcase $d_1,d_3\leq d_2$
is treated symmetrically.) Further,  we
give a construction of a statistical
operator $\rho_3$ having the given
dimensions.

We construct $\rho_3$ in spectral form
$$ \rho_3=\sum_{n=1}^{d_3}r_n
\ket{\Psi^{(n)}}_3\bra{\Psi^{(n)}}_3.$$
The characteristic values
$\{r_n:n=1,2,\dots ,d_3\}$ are
arbitrary fixed positive numbers such
that $\sum_{n=1}^{d_3}r_n=1$.

To construct the characteristic vectors
$\ket{\Psi^{(n)}}_3$, we introduce an
ON basis $\{\ket{i}_1:i=1,2,\dots
,d_1\}$ spanning $\cR(\rho_1)$, and
another ON basis
$\{\ket{j}_2:j=1,2,\dots ,d_2\}$
spanning $\cR(\rho_2)$. We break up the
former basis into $d_3$ disjoint sets
of basis vectors, i. e., into subbases,
each containing at most $d_2$ vectors.
(This is possible because of the first
inequality in (1).)  We enumerate the
subbases by $n=1,2,\dots ,d_3$. Let
$D_n$ be the number of basis vectors in
the $nth$ subbasis. Within this
subbasis we enumerate the vectors by a
subset of the indices of the chosen ON
basis in $\cR(\rho_2)$ as follows:
$$j=\Big(\sum_{n'=1}^{n-1}D_{n'}\Big)+1,
\enskip
\Big(\sum_{n'=1}^{n-1}D_{n'}\Big)
\enskip +2,\enskip \dots ,\enskip
\Big(\sum_{n'=1}^{n-1}D_{n'}\Big)+D_n.$$
(In the first subbasis the sums in
parentheses are, of course, omitted.)
When $j$ reaches the value $d_2$, we
count its values further cyclically:
$j=d_2+1\equiv 1,\enskip j=d_2+2\equiv
2$, etc.

Then we construct $$
\ket{\Psi^{(n)}}_3\equiv \sum_j\alpha_j
\ket{j}_1\otimes \ket{j}_2,\quad
n=1,2,\dots ,d_3,$$ where the
$\alpha_j$ are arbitrary nonzero
complex numbers such that
$\sum_j|\alpha_j|^2=1$ for each value
of $n$ independently, and "$j$"
enumerates the vectors $\ket{i}_1$
within the $n$th subbasis.

Obviously, on account of the
disjointness of the mentioned subbases,
$$ \bra{\Psi^{(n)}}\ket
{\Psi^{(n')}}=\delta_{n,n'}.$$

It is easily seen that the range
dimensions of the constructed $\rho_3$
are precisely the initially given
quantities $d_1,d_2,d_3$. This
completes the proof of claim (i).\\

{\it To prove claim} (ii), we first
prove {\it necessity} of the common
lower bound condition (2) {\it ab
contrario}.

{\it Lemma}. If $d_1=1$ or $d_2=1$ or
both, then the corresponding state
$\rho_3$ is necessarily uncorrelated,
i. e., $\rho_3 =\rho_1\otimes \rho_2$.

{\it Proof}. We assume $d_1=1$ (with no
assumption on $d_2$). Let $\ket{a}_1$
be a state vector spanning
$\cR(\rho_1)$, and let
$\{\ket{j}_2:j=1,2,\dots ,d_2\}$ be an
ON basis spanning $\cR(\rho_2)$. Since
$\cR(\rho_3)\subseteq
\Big(\cR(\rho_1)\otimes
\cR(\rho_2)\Big)$ (cf again \cite{FY},
relation (11) there), we can expand
$\rho_3$ in the diadic operator basis:
$$
\rho_3=\sum_j\sum_{j'}r_{jj'}\ket{a}_1
\bra{a}_1\otimes \ket{j}_2\bra{j'}_2=$$
$$\ket{a}_1 \bra{a}_1\otimes
\sum_j\sum_{j'}r_{jj'}\ket{j}_2\bra{j'}_2.$$
Taking the partial traces, one infers
that $\rho_1=\ket{a}_1 \bra{a}_1$, and
that $\rho_2=\sum_j\sum_{j'}
r_{jj'}\ket{j}_2\bra{j'}_2$, and,
finally, that $\rho_3=\rho_1\otimes
\rho_2$ as claimed. The case $d_2=1$
and $d_1\not= 1$ is proved
symmetrically. \hfill $\Box$\\

To prove {\it sufficiency} of condition
(2) in conjunction with (1) for claim
(ii), one should notice that both
constructions in the above proof of
claim (i) easily give a correlated
state in this case.\\

To {\it prove claim} (iii), we begin by
{\it necessity}. Let
$\rho_3=\rho_1\otimes \rho_2$, and let
$\{\ket{i}_1:i=1,2,\dots ,d_1\}$ and
$\{\ket{j}_2:j=1,2,\dots ,d_2\}$ be
characteristic subbases of $\rho_1$ and
$\rho_2$ respectively spanning the
respective ranges. Then $$ \rho_3=
\sum_{i=1}^{d_1}\sum_{j=1}^{d_2}r_ir_j
\ket{i}_1\bra{i}_1\otimes
\ket{j}_2\bra{j}_2,$$ where the $r_i$
and the $r_j$ are the corresponding
characteristic values of $\rho_1$ and
$\rho_2$ respectively. From this
characteristic decomposition of
$\rho_3$ with $d_1d_2$ terms one infers
that the product condition (3) is
valid.

To prove {\it sufficiency}, we
construct any $\rho_1$ and $\rho_2$
with the given range dimensions $d_1$
and $d_2$ respectively, and multiply
them: $\rho_3\equiv \rho_1\otimes
\rho_2$. \hfill $\Box$

\end{document}